\documentclass[10pt,conference]{IEEEtran}
\usepackage{pgfplots}
\IEEEoverridecommandlockouts
\usepackage{cite}
\usepackage{amsmath,amssymb,amsfonts,url}
\usepackage{graphicx}
\usepackage{textcomp}
\usepackage{xcolor}
\usetikzlibrary{pgfplots.groupplots}
\usepackage{comment}
\usepackage[nonumberlist]{glossaries}
\usepackage{multirow}
\usepackage[nonumberlist]{glossaries}
\usepackage{threeparttable}
\usepackage{graphicx} 
\usepackage{tikz}
\usetikzlibrary{arrows.meta}
\usetikzlibrary{positioning}
\pgfplotsset{compat=1.18}
\usepackage{subcaption}
\usepackage{acronym}
\usepackage{booktabs}
\usepackage{algorithm}
\usepackage{algpseudocode}
\usepackage[algo2e]{algorithm2e} 

\usepackage{xcolor} % Optional, only used for the gray comments
 \usepackage{float}
\def\BibTeX{{\rm B\kern-.05em{\sc i\kern-.025em b}\kern-.08em
    T\kern-.1667em\lower.7ex\hbox{E}\kern-.125emX}}

\newacronym{1D}{1D}{One Dimensional}
\newacronym{2D}{2D}{Two Dimensional}
\newacronym{6G}{6G}{Sixth Generation}
\newacronym{AI}{AI}{Artificial Intelligence}
\newacronym{BCE}{BCE}{Binary Cross-Entropy}
\newacronym{BER}{BER}{Bit Error Rate}
\newacronym{BLER}{BLER}{Block Error Rate}
\newacronym{CNN}{CNN}{Convolutional Neural Network}
\newacronym{CPU}{CPU}{Central Processing Unit}
\newacronym{CSI}{CSI}{Channel State Information}
\newacronym{DL}{DL}{Deep Learning}
\newacronym{DMRS}{DMRS}{Demodulation Reference Signal}
\newacronym{FFT}{FFT}{Fast Fourier Transform}
\newacronym{FLOP}{FLOP}{Floating-Point Operation}
\newacronym{GFLOP}{GFLOP}{Giga Floating-Point Operation}
\newacronym{FLOPs}{FLOPs}{Floating-Point Operations}
\newacronym{GPU}{GPU}{Graphics Processing Unit}
\newacronym{GNN}{GNN}{Graph Neural Network}
\newacronym{IFFT}{IFFT}{Inverse Fast Fourier Transform}
\newacronym{LN}{LN}{Layer Normalization}
\newacronym{LDPC}{LDPC}{Low-Density Parity-Check}
\newacronym{LLR}{LLR}{Log-Likelihood Ratio}
\newacronym{LLM}{LLM}{Large Language Model}
\newacronym{LMMSE}{LMMSE}{Linear Minimum Mean Squared Error}
\newacronym{LOS}{LOS}{Line-of-Sight}
\newacronym{LS}{LS}{Least-Squares}
\newacronym{ML}{ML}{Machine Learning}
\newacronym{MHSA}{MHSA}{Multi-Head Self-Attention }
\newacronym{MIMO}{MIMO}{Multiple-Input-Multiple-Output}
\newacronym{NLOS}{NLoS}{Non-LoS}
\newacronym{NPU}{NPU}{Neural Processing Unit}
\newacronym{NR}{NR}{New-Radio}
\newacronym{OFDM}{OFDM}{Orthogonal Frequency Division Multiplexing}
\newacronym{OOM}{OOM}{out-of-memory}
\newacronym{PHY}{PHY}{Physical Layer}
\newacronym{PTQ}{PTQ}{Post-Training Quantization}
\newacronym{QAM}{QAM}{Quadrature Amplitude Modulation}
\newacronym{QAT}{QAT}{Quantization-Aware Training}
\newacronym{RAN}{RAN}{Radio Access Network}
\newacronym{RB}{RB}{Resource Block}
\newacronym{RG}{RG}{Resource Grid}
\newacronym{SIMO}{SIMO}{Single-Input-Multiple-Output}
\newacronym{SISO}{SISO}{Single-Input-Single-Output}
\newacronym{SNR}{SNR}{Signal-to-Noise-Ratio}
\newacronym{STE}{STE}{Straight‑Through Estimator}
\newacronym{UE}{UE}{User Equipment}
\newacronym{URLLC}{URLLC}{Ultra-Reliable Low-Latency}
\newacronym{V2X}{V2X}{Vehicle-to-Everything}
\newacronym{ZF}{ZF}{Zero-Forcing}
\newacronym{PUSCH}{PUSCH}{Physical Uplink Shared Channel}
\newacronym{Q-DQ}{Q-DQ}{Quantize-Dequantize}
\newacronym{NMSE}{NMSE}{Normalized Mean Squared Error}
\newacronym{ReLU}{ReLU}{Rectified Linear Unit}
\newacronym{OCP}{OCP}{Open Compute Project}
\newacronym{ONNX}{ONNX}{Open Neural Network Exchange}
\newacronym{TDL}{TDL}{Tapped Delay Line}
\newacronym{MU-MIMO}{MU-MIMO}{Multi-User Multiple-Input Multiple-Output}
\newacronym{MCS}{MCS}{Modulation and Coding Scheme}
\newacronym{DNN}{DNN}{Deep Neural Network}
\newacronym{MSE}{MSE}{Mean Squared Error}
\newacronym{SRS}{SRS}{Sounding Reference Signal}
\newacronym{MAC}{MAC}{Multiply–accumulate}
\newacronym{PRB}{PRB}{Physical Resource Block}
\newacronym{CDM}{CDM}{Code Division Multiplexing}

\begin{document}

\title{{Performance Analysis of Quantization-Aware Neural Channel Estimation with Low-Precision Inference}}

\author{
    \IEEEauthorblockN{
    Joonas Törmälä\IEEEauthorrefmark{2},
    SaiKrishna Saketh Yellapragada\IEEEauthorrefmark{1},
    Eeli Susan\IEEEauthorrefmark{2},
    Esa Ollila\IEEEauthorrefmark{1}
    }
    \IEEEauthorblockA{
        \IEEEauthorrefmark{2}\textit{Nokia Solutions and Networks Oy, Finland},
        \IEEEauthorrefmark{1}\textit{Aalto University, Finland} \hspace{1em}
    }
}

\maketitle
\begin{abstract}
Neural channel estimators can improve radio performance of 5G receivers, but real-time deployment in \gls{RAN} imposes strict latency and power budgets. Prior studies of quantized neural channel estimators assess this cost through mean-squared error or analytic operation counts. In this paper, we train a 1D-\gls{CNN} channel estimator for the 5G \gls{NR} \gls{PUSCH} with INT8 and FP8 \gls{QAT}, and we evaluate its radio performance together with its latency, power, and energy on an NVIDIA H100 \gls{GPU}. On TDL-A and TDL-C channel at \gls{UE} speeds up to $35~\mathrm{m/s}$, both 8-bit models stay within $0.025~\mathrm{dB}$ of FP32 at $10\%$ \gls{BLER}. At batch size 128, INT8 reduces the latency by $16\%$, the incremental \gls{GPU} power by $33\%$, and the energy per query by $26\%$, compared with $6\%$, $26\%$, and $21\%$ for FP8. INT8 is therefore the preferred format.
\glsresetall
\end{abstract}
\begin{IEEEkeywords}
channel estimation, quantization-aware training, QAT, convolutional neural networks, 5G NR, PUSCH, INT8, FP8
\end{IEEEkeywords}

%\nocite{ho2019axial}
\section{Introduction}
\label{sec:intro}

Deep learning-based physical-layer processing is now a practical means of enhancing wireless receiver performance. Channel estimation is a key operation in the physical-layer receiver chain, since channel-estimate quality sets the reliability of the subsequent equalization and decoding. Conventional model-based estimators, including \gls{LS} and \gls{LMMSE} methods, remain the reference for \gls{OFDM} and for terrestrial 5G \gls{NR}~\cite{van1995channel,li1998robust,savaux2017lmmse}. Learned neural receivers and neural channel estimators improve on \gls{LS} and \gls{LMMSE} channel estimation in standard-compliant 5G \gls{NR}~\cite{deeprx,wiesmayr2025realtime,luan2023channelformer,saketh_axial}. Accuracy gains of learned neural receivers and neural channel estimators commonly come from architectures with a large computational and memory footprint, and a large footprint restricts deployment on platforms with tight latency and power limits.\par
Neural channel estimators learn the channel, and learned receivers can fold equalization and demapping into the same model, but real-time \glspl{RAN} imposes strict latency and power budgets. Reduced precision is one way to meet those budgets. The authors of~\cite{wiesmayr2025realtime} run \gls{MU-MIMO} \gls{PUSCH} processing that adapts to varying \glspl{MCS} without retraining, and achieve sub-millisecond inference with TensorRT on an NVIDIA A100 \gls{GPU} using FP16 weights. \gls{CNN} models retain near full-precision \gls{BLER} under 8-bit \gls{PTQ}, and \gls{QAT} extends this robustness down to 4-bit weights~\cite{saketh_asilomar25,saketh_qat}.\par
Architecture-level redesign also lowers the computational cost of learned baseband models. SpikingRx replaces artificial neurons in a \gls{CNN}-based model with spiking neurons, which approaches the link performance of the corresponding artificial-neuron model at an estimated $9\times$ lower energy, and remains robust to 8-bit weight quantization ~\cite{gupta2024spikingrx}. For channel estimation, the authors of~\cite{lee2023lightweight} proposed a single-hidden-layer \gls{DNN} that estimates the channel impulse response from \gls{SRS} pilots, and INT8 weights of the single-hidden-layer \gls{DNN} cause negligible \gls{MSE} degradation. CHEA replaces global attention with windowed multi-resolution attention for 5G \gls{NR} \gls{MU-MIMO} \gls{PUSCH} channel estimation~\cite{abdollahpour2026chea}. A single trained CHEA model supports different \gls{PRB} allocations, and its \gls{MAC} count scales linearly with the number of subcarriers. In addition, CHEA reports the lowest \gls{MSE} among the compared \gls{LS}, \gls{LMMSE}, convolutional, and attention-based estimators, at $2.8\times$ to $22.0\times$ fewer \glspl{MAC} than the attention-based estimators. Channelformer obtains a comparable parameter reduction by pruning an attention-based channel estimator, removing up to $70\%$ of the Channelformer parameters with little loss in estimation accuracy~\cite{luan2023channelformer}. Lightweight convolutional channel estimators for onboard inference report accuracy as \gls{MSE} and cost as a floating-point operation count~\cite{abdollahpour2026ntn_ce}.\par
No prior work has jointly quantified how INT8 and FP8 \gls{QAT} of a neural channel estimator affects \gls{BLER}, inference latency, power, and energy on a \gls{GPU}. We close that gap for a \gls{1D} \gls{CNN} estimator of the 5G \gls{NR} \gls{PUSCH}, trained in both formats and measured with TensorRT on an NVIDIA H100 \gls{GPU}. The comparison covers \gls{BLER} under \gls{UE} mobility, including a channel model withheld from training, together with measured latency, incremental \gls{GPU} power, and energy per query, and it uses those measurements to choose between the two 8-bit formats.

The rest of the paper is organized as follows. Section~\ref{sec:system_model} introduces the uplink model and \gls{QAT}. Section~\ref{sec:estimator} describes the quantized channel estimator. Section~\ref{sec:results} presents the link-level and hardware results, and Section~\ref{sec:conclusion} concludes the paper.

\section{System Model and Quantization-Aware Training}
\label{sec:system_model}

\subsection{Uplink Setup}
\label{sec:uplink}
We consider a 5G~\gls{NR} \gls{SIMO} uplink in which a single-antenna \gls{UE} transmits an \gls{OFDM} \gls{PUSCH} to a base station with $N_\text{Rx}$ receive antennas. The payload bits are encoded with a \gls{LDPC} code and mapped to \gls{QAM} symbols on a resource grid of $T$ \gls{OFDM} symbols and $F$ subcarriers, indexed by the symbol $n$ and the subcarrier $k$. A Type~1 \gls{DMRS} occupies every second subcarrier of two \gls{OFDM} symbols and provides $P$ pilots per \gls{DMRS} symbol.  
We denote the set of DMRS-symbol indices by $\mathcal{N}_\text{DMRS}$, with
$N_\text{DMRS}=|\mathcal{N}_\text{DMRS}|=2$, and the set of pilot-subcarrier indices in DMRS symbol $n$ by $\mathcal{P}_n$, with $|\mathcal{P}_n|=P$ for every $n\in\mathcal{N}_\text{DMRS}$.
The channel follows a 3GPP \gls{TDL} model at carrier frequency $f_c$. For a \gls{UE} speed $v$, the maximum Doppler shift is $f_D = v f_c / c$, where $c$ denotes the speed of light.

At the receiver, after cyclic-prefix removal and the \gls{FFT}, the received vector at symbol $n$ and subcarrier $k$ is
\begin{equation}
  \mathbf{y}_{n,k}
  = \mathbf{h}_{n,k}\, x_{n,k} + \mathbf{n}_{n,k},
\end{equation}
where $\mathbf{y}_{n,k}\in\mathbb{C}^{N_\text{Rx}\times 1}$ is the received signal vector, $\mathbf{h}_{n,k}\in\mathbb{C}^{N_\text{Rx}\times 1}$ is the channel frequency response, $x_{n,k}$ is the transmitted symbol with $\mathbb{E}[|x_{n,k}|^2]=1$, and $\mathbf{n}_{n,k}\sim\mathcal{CN}(\mathbf{0},\sigma^2\mathbf{I}_{N_\text{Rx}})$ is the noise. Here, $\mathcal{CN}$ denotes a circularly symmetric complex Gaussian distribution, and $\mathbf{I}_{N_\text{Rx}}$ is the $N_\text{Rx}$-dimensional identity matrix. The scalar channel coefficient on receive
antenna $r$ is denoted by $h_{r,n,k}$, so that $
  \mathbf{h}_{n,k}    = [h_{1,n,k},\ldots,h_{N_\text{Rx},n,k}]^{\top}.$
Similarly, $y_{r,n,k}$ denotes the $r$-th component of $\mathbf{y}_{n,k}$.
Table~\ref{tab:sim_params} summarizes the link, training, and evaluation parameters.

\begin{table}[h]
    \centering
    \caption{Simulation parameters.}
    \label{tab:sim_params}
    \begin{tabular}{ll}
        \hline
        Parameter & Value \\
        \hline
        \multicolumn{2}{l}{\textit{Link}} \\
        Receive antennas $N_\text{Rx}$ & 4 \\
        Subcarrier spacing & $15~\mathrm{kHz}$ \\
        \gls{OFDM} symbols $T$ & 14 \\
        Subcarriers $F$ & 96 (8~PRBs) \\
        \gls{DMRS} & Type~1, 2 symbols \\
        Pilots per \gls{DMRS} symbol $P$ & 48 \\
        Modulation and coding & 64-QAM (MCS~20) \\
        Carrier frequency $f_c$ & $3.5~\mathrm{GHz}$ \\
        Delay spread & $100~\mathrm{ns}$ \\
        \hline
        \multicolumn{2}{l}{\textit{Training}} \\
        Channel & TDL-A \\
        \gls{UE} speed $v$ & $\mathcal{U}(0, 35)~\mathrm{m/s}$ \\
        \gls{SNR} & $\mathcal{U}(-5, 20)~\mathrm{dB}$ \\
        Epochs $\times$ steps & $100 \times 200$ \\
        Optimizer & Adam \\
        Learning rate & Cosine, $10^{-3}$ to $10^{-5}$ \\
        \hline
        \multicolumn{2}{l}{\textit{Evaluation}} \\
        Channels & TDL-A, TDL-C \\
        \gls{UE} speed $v$ & 10, 20, $35~\mathrm{m/s}$ \\
        $E_b/N_0$ range & 6 to $7.5~\mathrm{dB}$, $0.1~\mathrm{dB}$ step \\
        Slots per $E_b/N_0$ point & 25,600 \\
        \hline
    \end{tabular}
\end{table}

\subsection{Channel Estimation}
\label{sec:channel_estimation}
At pilot subcarrier  $p\in\mathcal{P}_n$ of \gls{DMRS} symbol
$n\in\mathcal{N}_\text{DMRS}$, 
%$p$ of \gls{DMRS} symbol $n$, 
the \gls{LS} estimate on receive antenna $r$ is
\begin{equation}
\label{eq:ls_estimate}
  \hat{h}^{\text{LS}}_{r,n,p} = y_{r,n,p} / x_{n,p}.
\end{equation}
These estimates are noisy and available on only half of the subcarriers. The baseline \gls{LS} estimator of Sionna~\cite{sionna} averages each pair of adjacent pilot estimates, which separates the ports of a \gls{CDM} group, and then interpolates linearly in frequency and time to the full resource grid. Our channel estimator in Section~\ref{sec:estimator} replaces the averaging and the frequency interpolation: for each receive antenna $r$ and DMRS symbol $n$, it maps the $P$ pilot-based estimates $\{\hat{h}^{\text{LS}}_{r,n,p}:p\in\mathcal{P}_n\}$ to the full set of $F$ subcarrier estimates $\{\hat{h}_{r,n,k}:k=1,\ldots,F\}$.
%it maps the $P$ \gls{LS} estimates of each \gls{DMRS} symbol directly to all $F$ subcarriers, and 
Linear interpolation in time between the two \gls{DMRS} symbols provides the channel on the data symbols. Both receivers then apply \gls{LMMSE} equalization with the error variance of the averaged \gls{LS} estimate, soft demapping, and \gls{LDPC} decoding.

% ---- previous version ----
% These estimates are noisy and available on only half of the subcarriers. The channel estimator in Section~\ref{sec:estimator} therefore maps them to all $F$ subcarriers of each \gls{DMRS} symbol, and linear interpolation in time between the two \gls{DMRS} symbols provides the channel on the data symbols. \gls{LMMSE} equalization, soft demapping, and \gls{LDPC} decoding then follow the conventional \gls{PUSCH} receiver.
\begin{figure}[t]
    \centering
    \newcommand{\cnnchain}[1]{%
        \node[io,  minimum height=3.4mm]  (#1in)  at (0,0)   {};
        \node[blk, minimum height=6.2mm]  (#1c1)  at (1.1,0) {};
        \node[blk, minimum height=9.4mm]  (#1c2)  at (2.2,0) {};
        \node[blk, minimum height=15.8mm] (#1c3)  at (3.3,0) {};
        \node[blk, minimum height=9.4mm]  (#1c4)  at (4.4,0) {};
        \node[blk, minimum height=3.4mm]  (#1c5)  at (5.5,0) {};
        \node[fixed, minimum height=3.4mm] (#1up) at (6.6,0) {};
        \node[io,  minimum height=3.4mm]  (#1out) at (7.7,0) {};
    }
    \begin{tikzpicture}[
        font=\scriptsize,
        >={Stealth[length=1.2mm]},
        blk/.style={draw, fill=gray!20, minimum width=4mm, inner sep=0pt},
        io/.style={draw, fill=white, minimum width=4mm, inner sep=0pt},
        fixed/.style={draw, dashed, fill=white, minimum width=4mm, inner sep=0pt},
        qdq/.style={draw, circle, fill=white, minimum size=3mm, inner sep=0pt, font=\tiny},
    ]
        % (a) FP32 estimator
        \node[anchor=west, font=\footnotesize] at (-0.3,1.2) {(a) FP32};
        \cnnchain{a}
        \foreach \n/\c in {in/2, c1/16, c2/32, c3/64, c4/32, c5/2, up/2, out/2}
            \node[above=1pt of a\n] {\c};
        \foreach \s/\t in {in/c1, c1/c2, c2/c3, c3/c4, c4/c5, c5/up, up/out}
            \draw[->] (a\s) -- (a\t);
        \draw[decorate, decoration={brace, mirror, amplitude=3pt}]
            (0.9,-0.9) -- (4.6,-0.9) node[midway, below=3pt] {Conv1d ($k{=}5$) + ReLU};
        \node[align=center] at (0,-1.2)   {$\hat{\mathbf{h}}^{\text{LS}}_{r,n}$\\$48{\times}2$};
        \node[align=center] at (5.5,-1.2) {Conv1d\\$k{=}5$};
        \node[align=center] at (6.6,-1.2) {Lin.\\$48{\to}96$};
        \node[align=center] at (7.7,-1.2) {$\mathbf{a}_{r,n}$\\$96{\times}2$};

        % (b) QAT estimator with Q-DQ nodes
        \begin{scope}[yshift=-4.0cm]
            \node[anchor=west, font=\footnotesize] at (-0.3,2.1) {(b) INT8 / FP8 QAT};
            \cnnchain{b}
            \foreach \q/\x/\s/\t in {0/0.55/in/c1, 1/1.65/c1/c2, 2/2.75/c2/c3, 3/3.85/c3/c4, 4/4.95/c4/c5} {
                \node[qdq] (bq\q) at (\x,0) {Q};
                \draw (b\s) -- (bq\q);
                \draw[->] (bq\q) -- (b\t);
            }
            \foreach \s/\t in {c5/up, up/out}
                \draw[->] (b\s) -- (b\t);
            \foreach \c in {1,...,5} {
                \node[qdq] (bw\c) at ({1.1*\c},1.1) {Q};
                \node (bW\c) at ({1.1*\c},1.55) {$\mathbf{W}_{\c}$};
                \draw[->] (bW\c) -- (bw\c);
                \draw[->] (bw\c) -- (bc\c);
            }
            \node[above=1pt of bup, font=\tiny] {float};
            \node[above=1pt of bout, font=\tiny] {float};
            \node[align=center, text width=8cm] at (3.85,-1.25)
                {Q: quantize--dequantize. INT8: per-tensor weights and activations.\\
                 FP8 E4M3: per-channel weights, per-tensor activations.};
        \end{scope}
    \end{tikzpicture}
    \caption{1D \gls{CNN} channel estimator: (a) FP32 model and (b) \gls{QAT} model with \gls{Q-DQ} nodes on the input, the weights, and the \gls{ReLU} outputs. Block height grows with the number of output channels, indicated above each block. Every convolution keeps the length of 48 pilots, and the last convolution and the upsampling remain in floating point.}
    \label{fig:cnn_arch}
\end{figure}
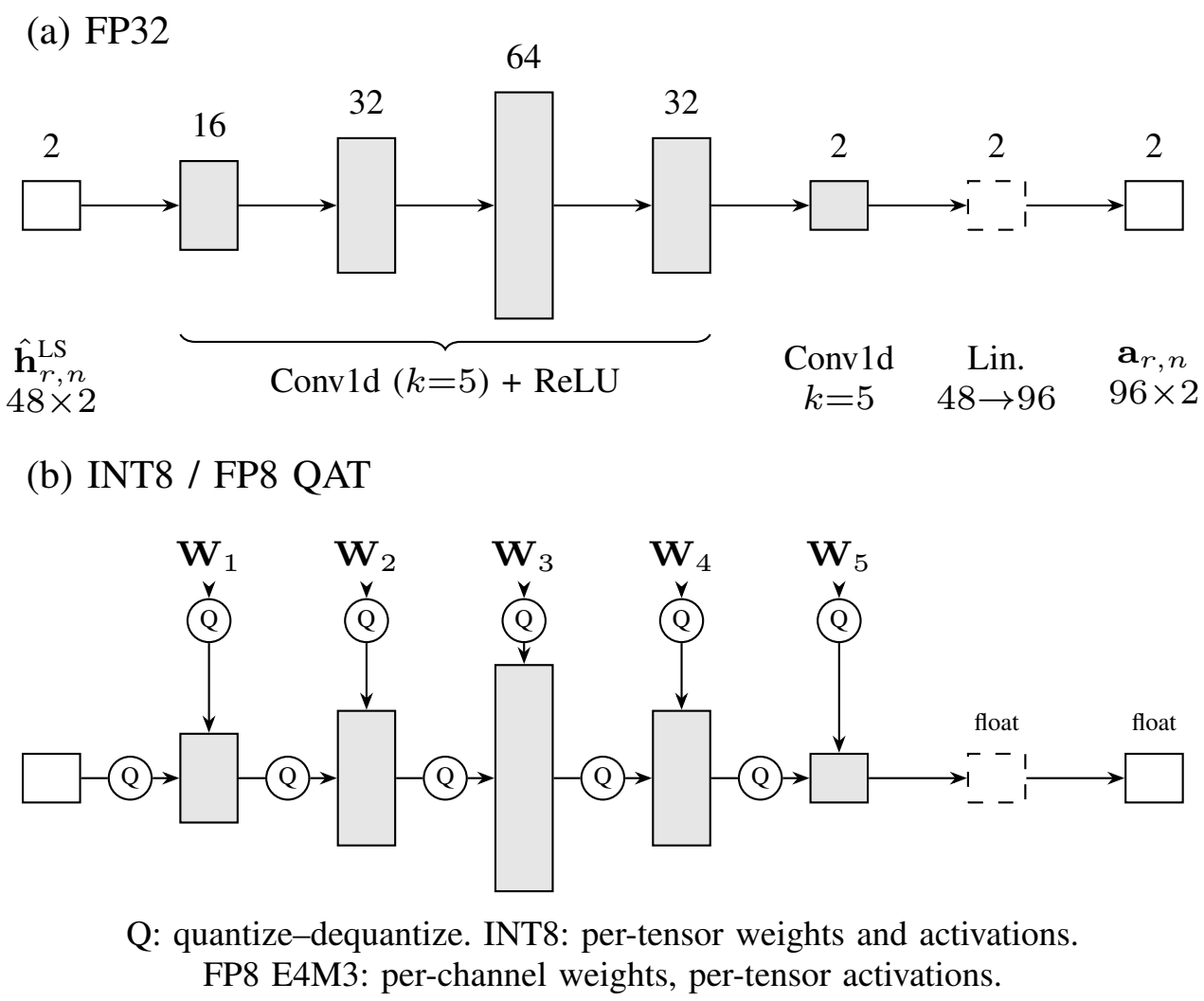

\begin{figure*}[t]
    \centering
    \includegraphics[width=\linewidth]{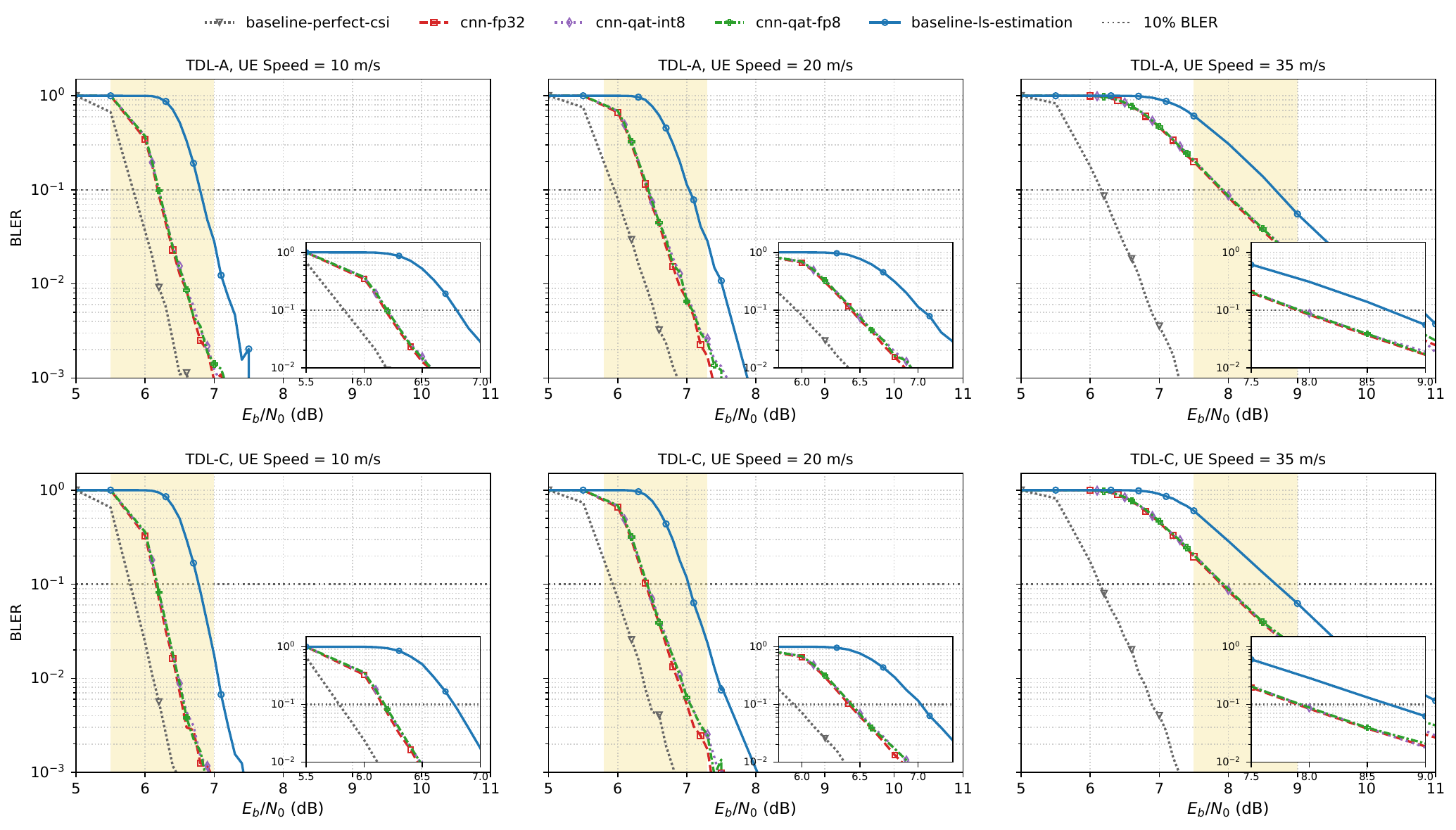}
    \caption{BLER versus $E_b/N_0$ for 64-QAM (MCS~20) with ideal \gls{CSI} and the baseline LS, FP32, INT8 and FP8 estimators. Rows show TDL-A and TDL-C, columns show \gls{UE} speeds of 10, 20 and $35~\mathrm{m/s}$, and the insets enlarge the shaded 5.5--$7~\mathrm{dB}$, 5.8--$7.3~\mathrm{dB}$ and 7.5--$9~\mathrm{dB}$ regions respectively.}
    \label{fig:bler_speed}
\end{figure*}

\subsection{Quantization-Aware Training}
\label{sec:qat}
\gls{QAT} simulates the target numerical format by inserting fake quantizers, implemented as \gls{Q-DQ} nodes, into the forward pass. Each node quantizes a tensor and immediately dequantizes it, so that the loss accounts for the rounding error of that format. For 8-bit integer quantization (INT8), each real-valued entry $x$ with scale $s$ is mapped to
\begin{equation}
\label{eq:fake_quant}
  \tilde{x}
  = s \cdot \operatorname{clip}\bigl(
      \lfloor x/s \rceil, q_{\min}, q_{\max}\bigr),
\end{equation}
where $\lfloor\cdot\rceil$ denotes rounding to the nearest integer and $[q_{\min}, q_{\max}]$ is the range of a signed or unsigned 8-bit integer. For 8-bit floating-point quantization (FP8), we use the \gls{OCP} E4M3 format, which allocates one sign bit, four exponent bits, and three mantissa bits. The fake quantizer rounds $x/s$ to the nearest E4M3 value and multiplies the result by $s$. The E4M3 levels are nonuniform and are densest near zero. The scale $s$ is either common to a tensor (per-tensor) or specific to each output channel of a weight tensor (per-channel); Section~\ref{sec:quantization} states which choice each format uses. In the backward pass, the \gls{STE}~\cite{bengio2013estimating} passes the gradient through the rounding operation unchanged, so the optimizer updates the full-precision weights.

\section{Quantized Channel Estimator}
\label{sec:estimator}

\subsection{Network Architecture}
\label{sec:architecture}
The estimator $f_{\boldsymbol{\theta}}$ processes each receive antenna
and each \gls{DMRS} symbol independently. 
%Its input $\mathbf{z}\in\mathbb{R}^{P\times 2}$ contains the real and imaginary parts of the $P$ \gls{LS} estimates in~(\ref{eq:ls_estimate}). 
For receive antenna $r$ and DMRS symbol $n\in\mathcal{N}_\text{DMRS}$, its input $\hat{\mathbf{h}}^{\text{LS}}_{r,n}\in\mathbb{R}^{P\times 2}$ contains the real and imaginary parts of the $P$ \gls{LS} estimates in~\eqref{eq:ls_estimate}.
As shown in Fig.~\ref{fig:cnn_arch}(a), the network consists of five 1D convolutional layers with kernel size $5$ and the channel sequence $2{\to}16{\to}32{\to}64{\to}32{\to}2$, and each of the first four layers is followed by a \gls{ReLU}. The upsampling from the pilot grid to the subcarrier grid has no trainable parameters, so we implement it as a fixed linear interpolation $\mathcal{U}$, applied
separately to the two output feature channels, that maps the $P$ outputs to the $F$ subcarriers,
% \begin{equation}
%   \hat{\mathbf{h}}
%   = \mathcal{U}\bigl(f_{\boldsymbol{\theta}}(\mathbf{z})\bigr)
%   \in \mathbb{C}^{F}.
% \end{equation}

\begin{equation}
  \mathbf{a}_{r,n}
  = \mathcal{U}\bigl(
      f_{\boldsymbol{\theta}}(\hat{\mathbf{h}}^{\text{LS}}_{r,n})
    \bigr)
  \in \mathbb{R}^{F\times 2}.
\end{equation}
The resulting complex channel estimate at subcarrier $k$ is
\begin{equation}
  \hat{h}_{r,n,k}
  = [\mathbf{a}_{r,n}]_{k,1}
    + \mathrm{j}[\mathbf{a}_{r,n}]_{k,2},
  \qquad k=1,\ldots,F,
\end{equation}
The model has $23{,}666$ parameters and requires $1.13\times10^{6}$ \gls{MAC} operations per receive antenna and \gls{DMRS} symbol, of which the two 64-channel layers account for $87\%$. One \gls{DMRS} symbol over all $N_\text{Rx}=4$ antennas thus requires $4.52\times10^{6}$ \glspl{MAC}, whereas the upsampling adds fewer than $10^{3}$ operations.

\subsection{INT8 and FP8 Quantization}
\label{sec:quantization}
For \gls{QAT}, Brevitas~\cite{brevitas} inserts the fake quantizers of Section~\ref{sec:qat} on the network input, the weights, and the \gls{ReLU} outputs (Fig.~\ref{fig:cnn_arch}(b)). INT8 uses per-tensor scales for weights and activations. For FP8, the weights use per-output-channel scales, so that the weights of each filter use the full dynamic range of E4M3, whereas the activations use per-tensor scales. Both quantized models are exported through \gls{ONNX} to TensorRT for inference.

\begin{figure}[t]
    \centering
    \includegraphics[width=0.8\columnwidth]{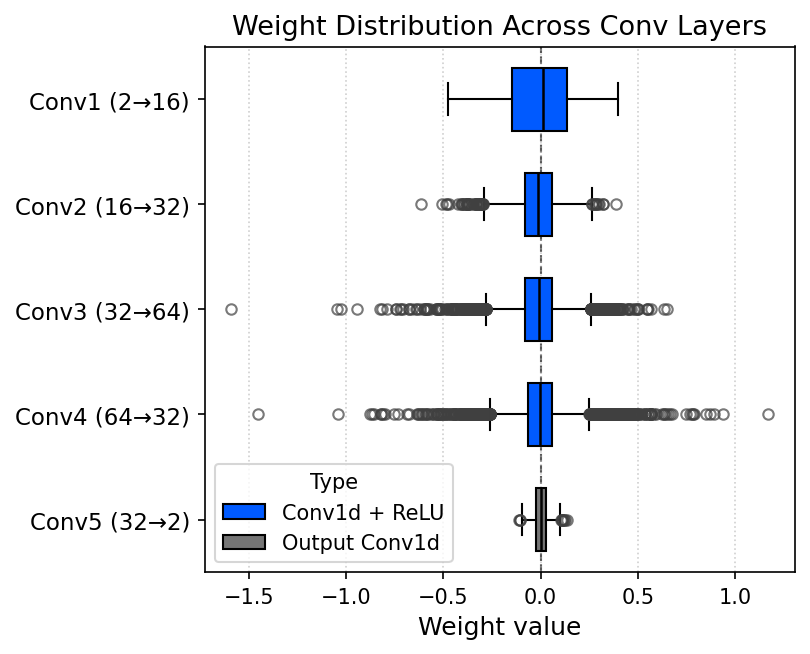}
    \caption{Weight distribution across the convolutional layers of the trained FP32 channel estimator. Boxes show the interquartile range, center lines the median, whiskers extend to $1.5$ times the interquartile range, and circles denote outliers. The central distributions are compact and centered near zero, while isolated outliers widen the ranges of Conv3 and Conv4. Such weight distributions motivate 8-bit \gls{QAT}, since INT8 per-tensor scaling and FP8 per-output-channel scaling can capture the layer-wise dynamic range with limited quantization error.}
    \label{fig:weight_distribution}
\end{figure}

The FP32 weight distributions further motivate this quantization strategy. As illustrated in Fig.~\ref{fig:weight_distribution}, all five convolutional layers are centered near zero. Conv2--Conv5 have compact central distributions, while the wider layer-wise ranges of Conv3 and Conv4 are caused by relatively few outliers. INT8 maps each layer's signed range to 8 bits with a per-tensor scale, whereas FP8 combines the exponent range of E4M3 with per-output-channel scaling. Both formats can therefore capture the observed dynamic range while retaining resolution around zero, which helps explain the small \gls{BLER} penalty reported in Section~\ref{sec:results}.

The estimator consists only of standard 1D convolutions, \gls{ReLU} activations, and linear interpolation, for which common deep learning frameworks and \gls{GPU} libraries provide optimized low-precision kernels~\cite{cudnn,rocm,miopen,pytorch,tensorflow,jax,tensorrt}. It can therefore be deployed without custom operators.
% The 1D convolution is a standard operator in cuDNN~\cite{cudnn}, ROCm and MIOpen~\cite{rocm,miopen}, PyTorch~\cite{pytorch}, TensorFlow~\cite{tensorflow}, JAX~\cite{jax}, and TensorRT~\cite{tensorrt}. As a result, \gls{GPU} and NPU toolchains map it onto optimized low-precision convolution kernels, and the estimator can be deployed without custom operators.

\subsection{Loss and Training Schedule}
\label{sec:training}
We train the estimator to minimize the per-subcarrier normalized squared error between the estimate and the true channel  coefficients 
%$\mathbf{h}$ 
of the \gls{DMRS} symbols. 
Let $h_{b,r,n,k}$ and
$\hat{h}_{b,r,n,k}$ denote the true and estimated
channel coefficients for slot $b$, receive antenna
$r$, OFDM symbol $n$, and subcarrier $k$. We define $N_\text{tr}=B N_\text{Rx}N_\text{DMRS}$ as the total number of receive-antenna and DMRS-symbol pairs across all slots in the batch.
The loss is  
\begin{equation}
\label{eq:nmse_loss}
  \mathcal{L}(\boldsymbol{\theta})
  = \frac{1}{N_\text{tr}F}
    \sum_{b=1}^{B}
    \sum_{r=1}^{N_\text{Rx}}
    \sum_{n\in\mathcal{N}_\text{DMRS}}
    \sum_{k=1}^{F}
    \frac{
      |\hat{h}_{b,r,n,k}-h_{b,r,n,k}|^2
    }{
      |h_{b,r,n,k}|^2+\epsilon
    }.
\end{equation}
% \begin{equation}
% \label{eq:nmse_loss}
%   \mathcal{L}(\boldsymbol{\theta})
%   = \frac{1}{N_\text{tr}F}
%     \sum_{i=1}^{N_\text{tr}}\sum_{k=1}^{F}
%     \frac{|\hat{h}_{i,k}-h_{i,k}|^2}{|h_{i,k}|^2+\epsilon},
% \end{equation}
where 
%$N_\text{tr}$ is the number of antenna and \gls{DMRS}-symbol pairs in a batch 
$\epsilon=10^{-7}$ stabilizes the denominator. The \gls{SNR} and \gls{UE} speed of each TDL-A training slot are drawn uniformly from the ranges in Table~\ref{tab:sim_params}. The FP32 model is trained with Adam and a cosine learning-rate schedule, %with 128 slots per step,  
using a batch size of $B=128$ slots, 
so that $N_\text{tr}=B N_\text{Rx}N_\text{DMRS} = 1024$. Both quantized models warm-start from the trained FP32 weights and are fine-tuned with the same schedule and the fake quantizers active.
\section{Experimental Results}
\label{sec:results}
We evaluate the FP32, INT8 and FP8 estimators in terms of \gls{BLER} and \gls{GPU} efficiency.

\textbf{Evaluation setup.} We simulate the \gls{BLER} in Sionna~\cite{sionna} on TDL-A and on TDL-C, which is not seen during training.

\textbf{Effect of \gls{UE} speed.} Fig.~\ref{fig:bler_speed} and Table~\ref{tab:snr10} show that the FP32 estimator operates close to ideal \gls{CSI} at low speed. At $10\%$ \gls{BLER}, the gap is $0.24~\mathrm{dB}$ at $10~\mathrm{m/s}$ and $0.45~\mathrm{dB}$ at $20~\mathrm{m/s}$. At $35~\mathrm{m/s}$, which corresponds to a maximum Doppler shift of $408~\mathrm{Hz}$ at $3.5~\mathrm{GHz}$, the gap grows to about $1.8~\mathrm{dB}$, and the \gls{BLER} remains near $1.4\times10^{-3}$ at $12~\mathrm{dB}$. This floor appears for all three precisions, so we attribute it to the pilot layout rather than to quantization. The \gls{CNN} refines each \gls{DMRS} symbol only in frequency, and linear interpolation between the two \gls{DMRS} symbols cannot follow the channel variation within the slot. The results on TDL-A and TDL-C differ by at most $0.03~\mathrm{dB}$.

\begin{table}[h]
    \centering
    \caption{$E_b/N_0$ (dB) at $10\%$ BLER and the penalty of the quantized models relative to FP32.}
    \label{tab:snr10}
    \begin{tabular}{lcccc}
        \hline
        Channel, speed & Ideal & FP32 & $\Delta$INT8 & $\Delta$FP8 \\
        \hline
        TDL-A, 10~m/s & 5.95 & 6.19 & +0.010 & +0.011 \\
        TDL-A, 20~m/s & 5.99 & 6.43 & +0.015 & +0.018 \\
        TDL-A, 35~m/s & 6.16 & 7.93 & +0.020 & +0.015 \\
        TDL-C, 10~m/s & 5.94 & 6.17 & +0.015 & +0.016 \\
        TDL-C, 20~m/s & 5.98 & 6.41 & +0.024 & +0.025 \\
        TDL-C, 35~m/s & 6.14 & 7.93 & +0.016 & +0.021 \\
        \hline
    \end{tabular}
\end{table}

\textbf{Effect of quantization.} As Table~\ref{tab:snr10} shows, INT8 and FP8 need 0.010--$0.025~\mathrm{dB}$ more $E_b/N_0$ than FP32 in all six cases. Between 6 and $7.5~\mathrm{dB}$, their \gls{BLER} is on average 1.10--1.33 times that of FP32 at 10 and $20~\mathrm{m/s}$. At $35~\mathrm{m/s}$, the ratio drops to 1.02, because mobility dominates the estimation error. The penalty is thus small, but it is consistent across both channels and all speeds. In contrast, INT8 and FP8 show no consistent ordering relative to each other.

\definecolor{fp32col}{HTML}{D62728}
\definecolor{int8col}{HTML}{9467BD}
\definecolor{fp8col}{HTML}{2CA02C}
\pgfplotsset{
    discard if not/.style 2 args={
        x filter/.append code={
            \edef\tempa{\thisrow{#1}}\edef\tempb{#2}%
            \ifx\tempa\tempb\else\def\pgfmathresult{inf}\fi
        }
    },
    fp32bar/.style={draw=fp32col, fill=fp32col!50, bar shift=0pt, discard if not={model}{fp32}},
    int8bar/.style={draw=int8col, fill=int8col!50, bar shift=0pt, discard if not={model}{int8}},
    fp8bar/.style={draw=fp8col, fill=fp8col!50, bar shift=0pt, discard if not={model}{fp8}},
}

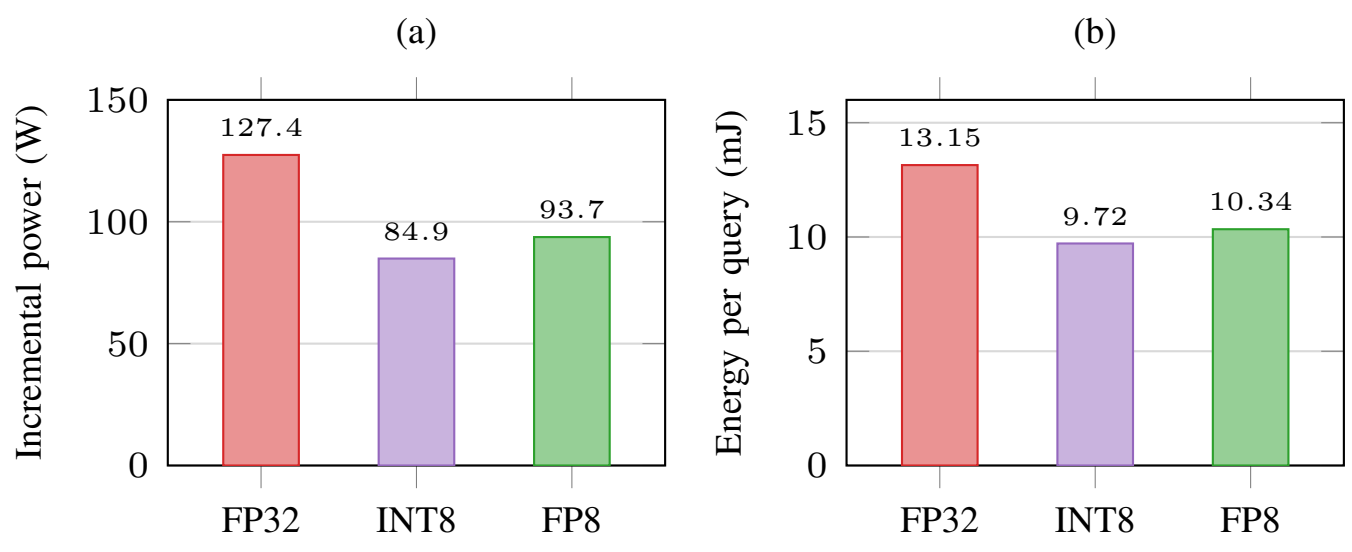
\begin{figure}[t]
    \centering
    \begin{tikzpicture}
        \begin{groupplot}[
            group style={group size=2 by 1, horizontal sep=1.2cm},
            every axis/.append style={
                width=0.55\columnwidth, height=4cm,
                ybar, bar width=5mm, ymin=0, enlarge x limits=0.3,
                symbolic x coords={fp32,int8,fp8}, xtick={fp32,int8,fp8},
                xticklabels={FP32,INT8,FP8},
                tick label style={font=\scriptsize},
                label style={font=\scriptsize},
                title style={font=\scriptsize},
                every node near coord/.append style={font=\tiny, text=black},
                ymajorgrids, grid style={gray!30},
                table/col sep=comma, unbounded coords=discard,
            },
        ]
            \nextgroupplot[ylabel={Incremental power (W)}, ymax=150, title={(a)},
                nodes near coords={\pgfmathprintnumber[fixed, fixed zerofill, precision=1]{\pgfplotspointmeta}}]
            \addplot[fp32bar] table[x=model, y=delta_power_mean_w] {figures/power_b128.csv};
            \addplot[int8bar] table[x=model, y=delta_power_mean_w] {figures/power_b128.csv};
            \addplot[fp8bar]  table[x=model, y=delta_power_mean_w] {figures/power_b128.csv};
            \nextgroupplot[ylabel={Energy per query (mJ)}, ymax=16, title={(b)},
                nodes near coords={\pgfmathprintnumber[fixed, fixed zerofill, precision=2]{\pgfplotspointmeta}}]
            \addplot[fp32bar] table[x=model, y=energy_per_inference_mj] {figures/power_b128.csv};
            \addplot[int8bar] table[x=model, y=energy_per_inference_mj] {figures/power_b128.csv};
            \addplot[fp8bar]  table[x=model, y=energy_per_inference_mj] {figures/power_b128.csv};
        \end{groupplot}
    \end{tikzpicture}
    \caption{GPU efficiency on an NVIDIA H100 with TensorRT at batch size 128. (a) Mean power above the $70.1~\mathrm{W}$ idle level. (b) Energy per query.}
    \label{fig:power_energy}
\end{figure}

\begin{figure}[t]
    \centering
    \begin{tikzpicture}
        \begin{axis}[
            width=0.9\columnwidth, height=6.5cm, font=\scriptsize,
            xmode=log, log basis x=2, ymode=log,
            xlabel={Batch size}, ylabel={Mean latency (ms)},
            xmin=1, xmax=16384, grid=both, grid style={gray!30},
            legend pos=north west, legend cell align=left,
            table/col sep=comma, unbounded coords=discard,
        ]
            \addplot[fp32col, solid, mark=o, mark size=1.2pt, discard if not={model}{fp32}]
                table[x=batch_size, y=latency_mean_ms] {figures/benchmark_results.csv};
            \addplot[int8col, dashed, mark=square, mark size=1.2pt, discard if not={model}{int8}]
                table[x=batch_size, y=latency_mean_ms] {figures/benchmark_results.csv};
            \addplot[fp8col, dash dot, mark=triangle, mark size=1.2pt, discard if not={model}{fp8}]
                table[x=batch_size, y=latency_mean_ms] {figures/benchmark_results.csv};
            \legend{FP32, INT8, FP8}
        \end{axis}
    \end{tikzpicture}
    \caption{Mean TensorRT latency on an NVIDIA H100 versus batch size for the FP32, INT8 and FP8 estimators.}
    \label{fig:latency}
\end{figure}
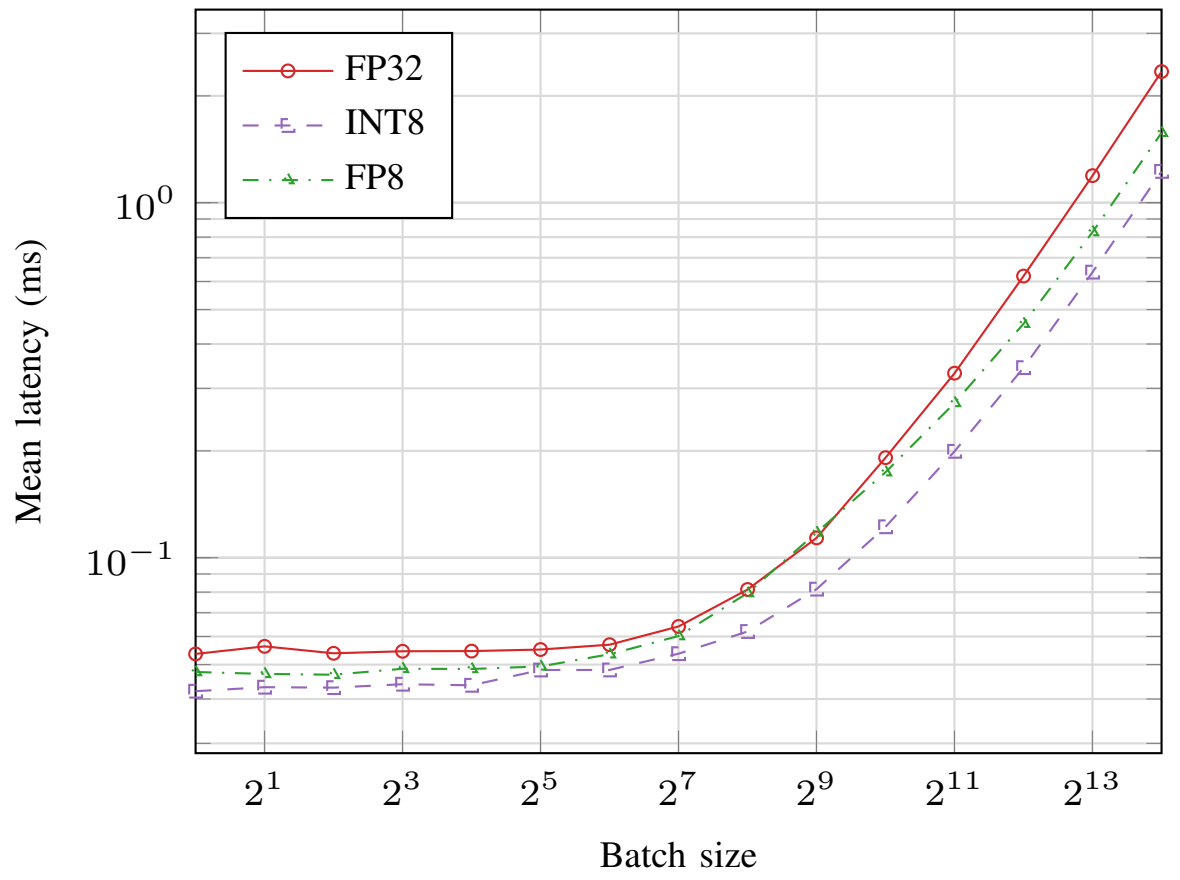

\textbf{Hardware efficiency.} Fig.~\ref{fig:latency} shows the mean TensorRT latency on an NVIDIA H100 as a function of batch size. At batch size of 128, INT8 reduces the latency by $16\%$, from $0.064$ to $0.054~\mathrm{ms}$, whereas FP8 reduces it by $6\%$. Power and energy follow the same order (Fig.~\ref{fig:power_energy}). Above the idle level of $70.1~\mathrm{W}$, the \gls{GPU} draws additional $127.4~\mathrm{W}$ with FP32, $84.9~\mathrm{W}$ with INT8 and $93.7~\mathrm{W}$ with FP8. The energy per query decreases accordingly, from $13.15~\mathrm{mJ}$ to $9.72~\mathrm{mJ}$ and $10.34~\mathrm{mJ}$. The reduction is larger at large batches: at batch size 16,384, INT8 reduces the \gls{GPU} compute time by $57\%$ and FP8 by $40\%$. Since both formats give the same radio performance, we consider INT8 the more efficient format.
%, in agreement with van Baalen et al.~\cite{vanbaalen2023fp8}.
\section{Conclusion and Future Work}
\label{sec:conclusion}
In this work, we evaluated quantization-aware training of a 1D \gls{CNN} channel estimator for 64-QAM and \gls{UE} speeds up to $35~\mathrm{m/s}$. Eight-bit \gls{QAT} costs at most $0.025~\mathrm{dB}$ at $10\%$ \gls{BLER} on both TDL-A and the unseen TDL-C, and we observe no consistent difference between INT8 and FP8. On an H100, INT8 reduces the incremental \gls{GPU} power by $33\%$ and the energy per query by $26\%$ relative to FP32, whereas FP8 reduces them by $26\%$ and $21\%$. INT8 is therefore the preferred format. At $35~\mathrm{m/s}$, all precisions lose about $1.8~\mathrm{dB}$ to ideal \gls{CSI}, because two \gls{DMRS} symbols with linear time interpolation cannot track the channel within the slot. Recent works on neural receivers have shown that FP4 microformat variants can outperform classical full-precision approaches with pruning and quantization \cite{yellapragada2026fpmicroquant}. Further optimization of the neural channel estimator using microformat approaches can be considered as future work.

% Microformat quantization and pruning compress learned baseband weights beyond 4-bit \gls{QAT}~\cite{saketh_qat,yellapragada2026fpmicroquant}, and the reported efficiency is \gls{BLER} or an operation count.

% This work proposes axial attention as a computationally efficient framework for neural receivers in AI-native 6G systems. By factorizing self-attention along temporal and spectral dimensions, the architecture overcomes the quadratic scalability bottleneck of conventional transformers while retaining the global context of the \gls{RG}. Our results demonstrate that the axial receiver achieves consistent performance gains over CNN and LS baselines, particularly at stringent 1\% BLER targets while reducing inference GFLOPs by over $3.5\times$ compared to CNNs. These properties make axial attention based architectures excellent candidate for resource-constrained edge deployments requiring ultra-reliable low-latency processing.
% Future work on axial attention architecture will focus on two key directions, namely extension to MIMO configurations and low-bit quantization.
%\input{sections/06_Conclusion}
%\newpage
\subsection*{Acknowledgment}
%The second author's work was supported by the European Union's 6GARROW project
%(No.~101192194).
The work of the second author has been supported by
the 6GARROW project which has received funding from the
Smart Networks and Services Joint Undertaking (SNS JU)
under the European Union’s Horizon Europe research and
innovation programme under Grant Agreement No 101192194
and from the Institute for Information \& Communications
Technology Promotion (IITP) grant funded by the Korean
government (MSIT) (No. RS-2024-00435652).
% arXiv does not always run BibTeX; ship \texttt{main.bbl} and use \texttt{\\input}.
% Regenerate with: pdflatex main && bibtex main && pdflatex main (twice).
\bibliographystyle{IEEEbib}

% \bibliography{references}

\end{document}